\documentclass{article}
\usepackage{waspaa17,amsmath,graphicx,url,times}
\usepackage{color}
\usepackage{enumitem}
\usepackage{subfig}

\title{Sound Event Detection in Synthetic Audio:\\ Analysis of the DCASE 2016 Task Results}

\name{Gr\'{e}goire Lafay$^1$, Emmanouil Benetos$^2$ and Mathieu Lagrange$^1$ \thanks{EB is supported by a UK RAEng Research Fellowship (RF/128).}}
\address{$^1$ LS2N, CNRS, Ecole Centrale de Nantes, France\\
$^2$ School of EECS, Queen Mary University of London, UK}

\begin{document}

\ninept
\maketitle

\begin{sloppy}

\begin{abstract}
As part of the 2016 public evaluation challenge on Detection and Classification of Acoustic Scenes and Events (DCASE 2016), the second task focused on evaluating sound event detection systems using synthetic mixtures of office sounds. This task, which follows the `Event Detection - Office Synthetic' task of DCASE 2013, studies the behaviour of tested algorithms when facing controlled levels of audio complexity with respect to background noise and polyphony/density, with the added benefit of a very accurate ground truth. This paper presents the task formulation, evaluation metrics, submitted systems, and provides a statistical analysis of the results achieved, with respect to various aspects of the evaluation dataset.
\end{abstract}

\begin{keywords}
Sound event detection, experimental validation, DCASE, acoustic scene analysis, sound scene analysis
\end{keywords}

\section{Introduction}

The emerging area of computational sound scene analysis (also called acoustic scene analysis) addresses the problem of automatically analysing environmental sounds, such as urban or nature sound environments \cite{Stowell15}. Applications of sound scene analysis include smart homes/cities, security/surveillance, ambient assisted living, sound indexing, and acoustic ecology. A core problem in sound scene analysis is sound event detection (SED), which involves detecting and classifying sounds of interest present in an audio recording, specified by start/end times and sound labels. A key challenge in SED is the detection of multiple overlapping sounds, also called \emph{polyphonic sound event detection}. To that end, the public evaluation challenge on Detection and Classification of Acoustic Scenes and Events (DCASE 2013) \cite{Stowell15} ran tasks for both monophonic and polyphonic SED, attracting 7 and 3 submissions respectively.

The follow-up challenge, DCASE 2016 \cite{DCASE2016}, attracted submissions from over 80 international teams amongst 4 different tasks.
The 2nd task of DCASE 2016 \cite{DCASE2016} is entitled ``Sound event detection in synthetic audio'', and follows the polyphonic ``Event Detection - Office Synthetic''  task of DCASE 2013. For the 2016 challenge, Task 2 attracted 10 submissions from 9 international teams, totalling 37 authors. The main aim of this task is to evaluate the performance of polyphonic sound event detection systems in office environments, using audio sequences generated by artificially concatenating isolated office sounds. The evaluation data was generated using varying levels of noise and polyphony. The benefits of this approach include very accurate ground truth, and the fact that the scene complexity can be broken into different types, allowing us to finely study the sensitivity of the evaluated algorithms to a given scene characteristic. Here, we evaluate differences between systems with respect to aspects of the dataset, including the event to background ratio (EBR), the scene type (monophonic vs. polyphonic scenes), and scene density. This paper presents the formulation of DCASE 2016 - Task 2, including generated datasets, evaluation metrics, submitted systems, as well as a detailed statistical analysis of the results achieved, with respect to various aspects of the evaluation dataset.


\section{Dataset Creation} \label{sec:datasets}

\subsection{Isolated Sounds} \label{sec:isolated}

The isolated sound dataset used for simulating the scenes has been recorded in the offices of LS2N (France). 11 sound classes have been considered, presented in Table ~\ref{tab:eventDCASE2016}. Compared to DCASE 2013, 5 classes are no longer considered: \textit{alert} (due to its too broad definition and large diversity compared to other classes), \textit{printer}, \textit{switch}, \textit{mouse click}, and \textit{pen drop}. After detailed post-analysis of the previous challenge, those latter classes that have been badly recognised and often confused were found to be too short in terms of duration. The class \textit{printer} is no longer considered since it is composed of sounds that are much longer than the ones of the other classes.

\subsection{Simulation process} \label{sec:ch7_simulationDcase2016}

For simulating the recordings used for Task 2, the morphological sound scene model proposed in \cite{Lafay16} is used.
Two core parameters are considered to simulate the acoustic scenes:
\begin{itemize}
\item $EBR$: the event-to-background ratio, i.e. the average amplitude ratio between sound events and the background
\item $nec$: the number of events per class
\end{itemize}

For each scene, the onset locations are randomly sampled using a uniform distribution. For monophonic scenes, a post-treatment that takes into account the duration of the event is done by moving some onsets in order to fulfill the constraint that two events must not overlap. 3 levels are considered for each parameter:  $EBR$: -6, 0 and +6~$dB$; $nec$: 1, 2, and 3 for a monophonic scene, and 3, 4, and 5 for polyphonic ones. This set of parameters leads to 18 experimental conditions.

\begin{table}
\begin{center}
\resizebox{200pt}{!}{
\begin{tabular}{lll}
\textbf{Index} & \textbf{Name}  & \textbf{Description}  \\
\hline
1   & door knock & knocking a door \\
2   & door slam & closing a door \\
3   & speech        & sentence of human speech \\
4   & laugh          & human laughter  \\
5   & throat         & human clearing throat  \\
6   & cough          & human coughing \\
7   & drawer        & open / close of a desk drawer \\
8   & keyboard       & computer keyboard typing \\
9   & keys         & dropping keys on a desk \\
10  & phone     & phone ringing \\
11  & page turn          & turning a page \\
\hline
\end{tabular}
}
\end{center}
\vspace{-0.1in}
\caption{Sound event classes used in DCASE 2016 - Task 2.}
\label{tab:eventDCASE2016}
\end{table}

\subsection{Task Datasets}

The training dataset consists of 220 isolated sounds containing 20 sounds per event class (see subsection \ref{sec:isolated}). The scenes of the development dataset are built using the isolated sounds of the training dataset mixed with 1 background sample.
One scene is built for each experimental condition, leading to 18 acoustic scenes. Only the $EBR$ is an independent parameter, so for each combination of $nec$ and monophony/polyphony, the chosen samples are different.

The scenes of the test dataset are built using the same rules from a set of 440 event samples, using 40 samples per class, and 3 background samples. For each of the 18 experimental conditions, the simulation is replicated 3 times, leading to a dataset of 54 acoustic scenes, with each scene being 2 minutes long. For each replication, the onset locations are different.
The total duration of the development and test datasets is 34 and 108 minutes, respectively.

\section{Evaluation} \label{sec:evaluation}

\subsection{Metrics}

Among the four metrics considered in the DCASE 2016 challenge \cite{Mesaros:2016}, the class-wise event-based F-measure ($F_{\mathit{cweb}}$) is considered in this paper for providing a statistical analysis of the results. This metric, which was also used in DCASE 2013 \cite{Stowell15} and on evaluating the scene synthesizer used in this work \cite{Lafay16}, assumes a correctly identified sound event if its onset is within 200~ms from a ground truth onset with the same class label. Results are computed per sound event class and are subsequently averaged amongst all classes.

\subsection{Statistical Analysis}

Contrary to the data published in the DCASE 2016 Challenge website \cite{DCASE2016} where the computation of the performance is done by concatenating all acoustic scenes along time and computing a single metric,
here performance is computed separately for each acoustic scene of the evaluation corpus, and averaged for specific groups of recordings, allowing us to perform statistical significance analyses.

Once computed, analysis of the results is done in 3 steps. First, results are considered globally, without considering the different experimental conditions individually. Differences between systems are analysed using a repeated measure ANOVA \cite{demvsar2006statistical} with one factor being the different systems. Second, the performance difference between monophonic and polyphonic acoustic scenes is analysed using a repeated measure ANOVA with one within-subject factor being the different systems and one between-subject factor being the monophonic/polyphonic factor. Third, the impact of $nec$ and $EBR$ on the performance of the systems considering separately the monophonic and polyphonic acoustic scenes is evaluated. For $nec$, the differences between the systems are evaluated using a repeated measure ANOVA with one within-subject factor being the different systems and one between-subject factor being $nec$. For the $EBR$ factor, performance differences between systems are analysed using a repeated measure ANOVA with 2 within-subject factors, namely the different systems and the $EBR$.

For the repeated measure ANOVA, the sphericity is evaluated with a Maulchy test \cite{mauchly1940}. If the sphericity is violated, the $p$-value is computed with a Greenhouse-Geisser correction \cite{Greenhouse1959}. In this case, we note $p_{gg}$ the corrected $p$-value. \emph{Post hoc} analysis is done by following the Tukey-Kramer procedure \cite{Tukey1949}. A significance threshold of $\alpha=0.05$ is chosen.

\section{Systems} \label{sec:systems}

\begin{table}[t]
\begin{center}
\resizebox{240pt}{!}{
\begin{tabular}{lcccc}
\hline
\textbf{System}             & \textbf{Features}         & \textbf{Classifier} &   \multicolumn{2}{c}{\textbf{Background}} \\
                             &                              &                      & \textbf{Reduction} & \textbf{Estimation}  \\
\hline \hline
\emph{Komatsu}   \cite{Komatsu2016}             &     VQT                      & NMF-MLD              &           & x \\
\hline
\emph{Choi}  \cite{Choi2016}                  &     Mel                      & DNN                  & x         & x \\
\hline
\emph{Hayashi 1}   \cite{Hayashi2016}           &     Mel                      & BLSTM-PP             &           & x \\
\hline
\emph{Hayashi 2} \cite{Hayashi2016}            &     Mel                      & BLSTM-HMM            &           & x \\
\hline
\emph{Phan}   \cite{Phan2016}                &     GTCC                     & RF                   &           & x\\
\hline
\emph{Giannoulis} \cite{Giannoulis2016}            &     Mel                      & CNMF                 &           & x\\
\hline
\emph{Pikrakis}   \cite{Pikrakis2016}             &     Bark                     & Template matching             & x         & \\
\hline
\emph{Vu}   \cite{Vu2016}                   &     CQT                      & RNN                  &           & \\
\hline
\emph{Gutierrez}  \cite{GutierrezArriola2016}            &     MFCC                     & KNN                  &           & x \\
\hline
\emph{Kong}   \cite{Kong2016}                 &     Mel                      & DNN                  &           &  \\
\hline
\emph{Baseline}  \cite{Benetos2016}              &     VQT                      & NMF                  &           &  \\
\hline
\end{tabular}
}
\end{center}
\vspace{-0.1in}
\caption{DCASE 2016 - Task 2: description of submitted systems. For brevity, acronyms are defined in the respective citations.}
\label{tab:systemsDcase2016}
\end{table}

For Task 2, 10 systems were submitted from various international research laboratories together with one \emph{baseline} system provided by the authors. An outline of the evaluated systems is provided in Table~\ref{tab:systemsDcase2016}. Most systems can be split into several successive processing blocks: a feature computation step optionally preceded by a denoising step (either estimating the background level in a training stage or performing background reduction) and a classification step where for each feature frame one or several event labels may be triggered.

Considering feature extraction, several design options have been taken, which are summarised below:
\begin{itemize}
\item \emph{mel/bark}: auditory model-inspired time/frequency representations with a non linear frequency axis
\item \emph{VQT/CQT}: variable- or constant-Q transforms
\item \emph{MFCC}: Mel-frequency cepstral coefficients
\item \emph{GTCC}: Gammatone filterbank cepstral coefficients
\end{itemize}

Several classifiers have been considered to process those features, ranging from nearest neighbour classifiers to spectrogram factorisation methods (such as NMF) to deep neural networks (such as RNNs). Classifiers are outlined in Table~\ref{tab:systemsDcase2016}, with more details to be found in the respective technical report of each submission \cite{Komatsu2016}-\cite{Benetos2016}.

\section{Results} \label{sec:results}

\subsection{Global Analysis} \label{sec:ch7_analyseGlobaleDcase2016}

\begin{figure}[t]
\includegraphics[width=1\columnwidth]{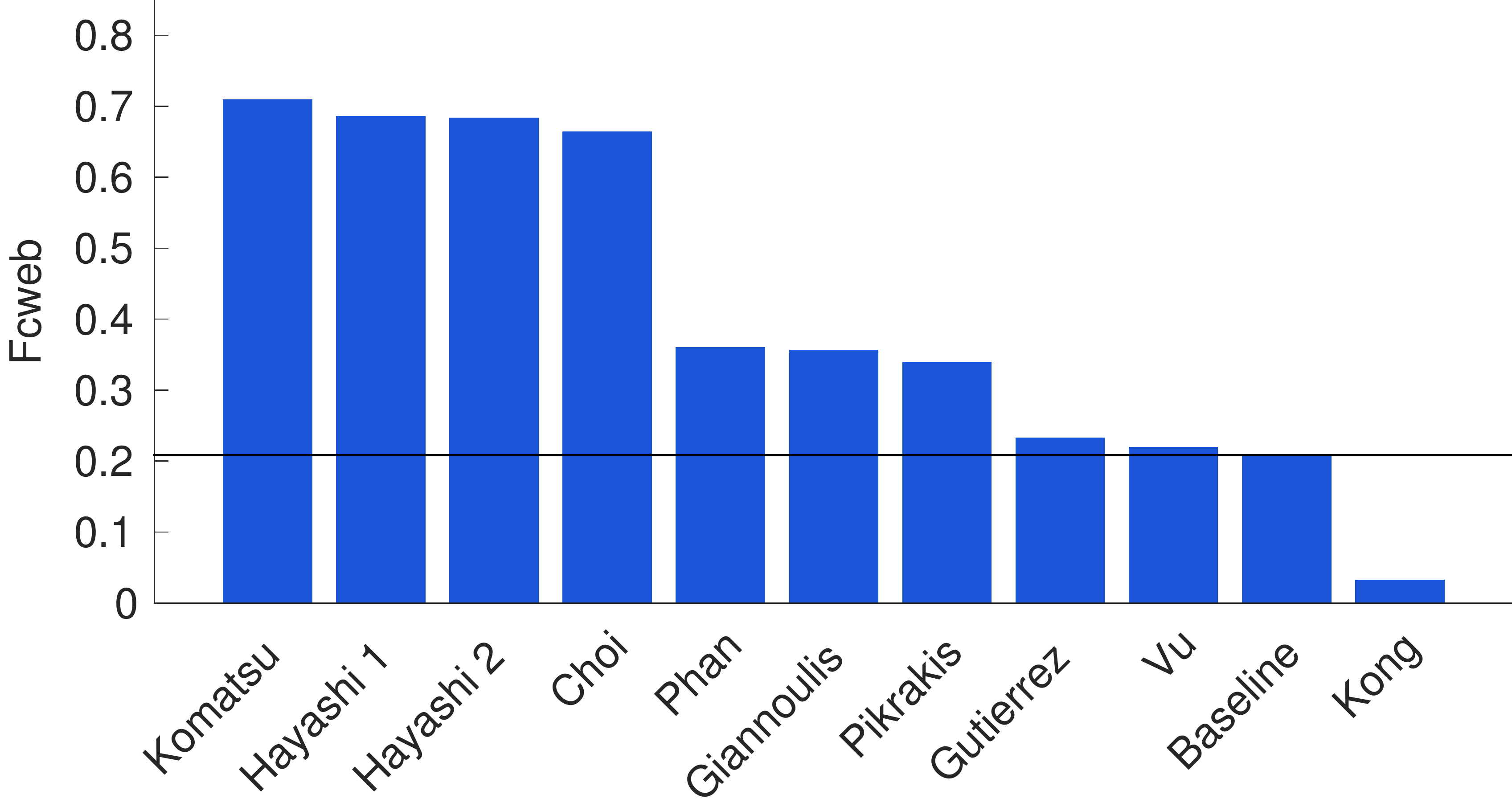}
\caption{Global performance of the systems submitted to DCASE 2016 - Task 2, in terms of $F_{\mathit{cweb}}$. Baseline performance is shown with a straight line.}
\vspace{-0.05in}
\label{fig:overall_eb_class_wise_F}
\end{figure}

Results in terms of $F_{\mathit{cweb}}$ are plotted on Fig.~\ref{fig:overall_eb_class_wise_F}. An ANOVA analysis on $F_{\mathit{cweb}}$ shows a positive effect of the system type: $F[10,530]=466$, $p_{gg}<0.01$ ($F$ stands for the F-statistic and its arguments stand for the degrees of freedom for systems and error, respectively).  \emph{Post hoc} analysis clusters systems into 4 groups where each pair of systems within a group does not have any statistical differences in terms of performance:
\begin{enumerate}[itemsep=-0.2mm]
\item \emph{Komatsu}, \emph{Hayashi 1}, \emph{Hayashi 2} and \emph{Choi}: performance ranges from 67\% (\emph{Choi}) to 71\% (\emph{Komatsu});
\item \emph{Phan}, \emph{Giannoulis} and \emph{Pikrakis}: performance ranges from 34\% (\emph{Pikrakis}) to 36\% (\emph{Phan});
\item \emph{Baseline}, \emph{Vu} and \emph{Gutierrez}: performance ranges from 21\% (\emph{Baseline}) to 23\% (\emph{Gutierrez});
\item \emph{Kong}: performance is 2\%.
\end{enumerate}

Among the 10 systems, 7 significantly improve upon the \emph{Baseline}. Systems of Group 2 improve around 15\%, while those of Group 1 reach an improvement close to 45\%. The impact of the chosen classifier is difficult to grasp from the analysis of the results, as there is a wide variety of chosen architecture for the systems of Group 1. Concerning features though, 3 among 4 use Mel spectrograms. Background estimation/reduction is of importance, as the 3 systems that do not explicitly handle it have the lowest performance.

The least performing system is \emph{Kong} as its results are systematically below the baseline. One possible explanation of this weak performance is the poor management of the training phase of the DNN classifier \cite{Kong2016}. It is known that such architecture requires a large amount of data in order to be robustly trained, and without any data augmentation strategy such as the one used by \emph{Choi} \cite{Choi2016}, the amount of data provided with the training dataset is not sufficient. The performance of \emph{Kong} will therefore not be discussed further.

\subsection{Monophonic vs. Polyphonic Scenes}

\begin{figure}[t]
\includegraphics[width=1\columnwidth]{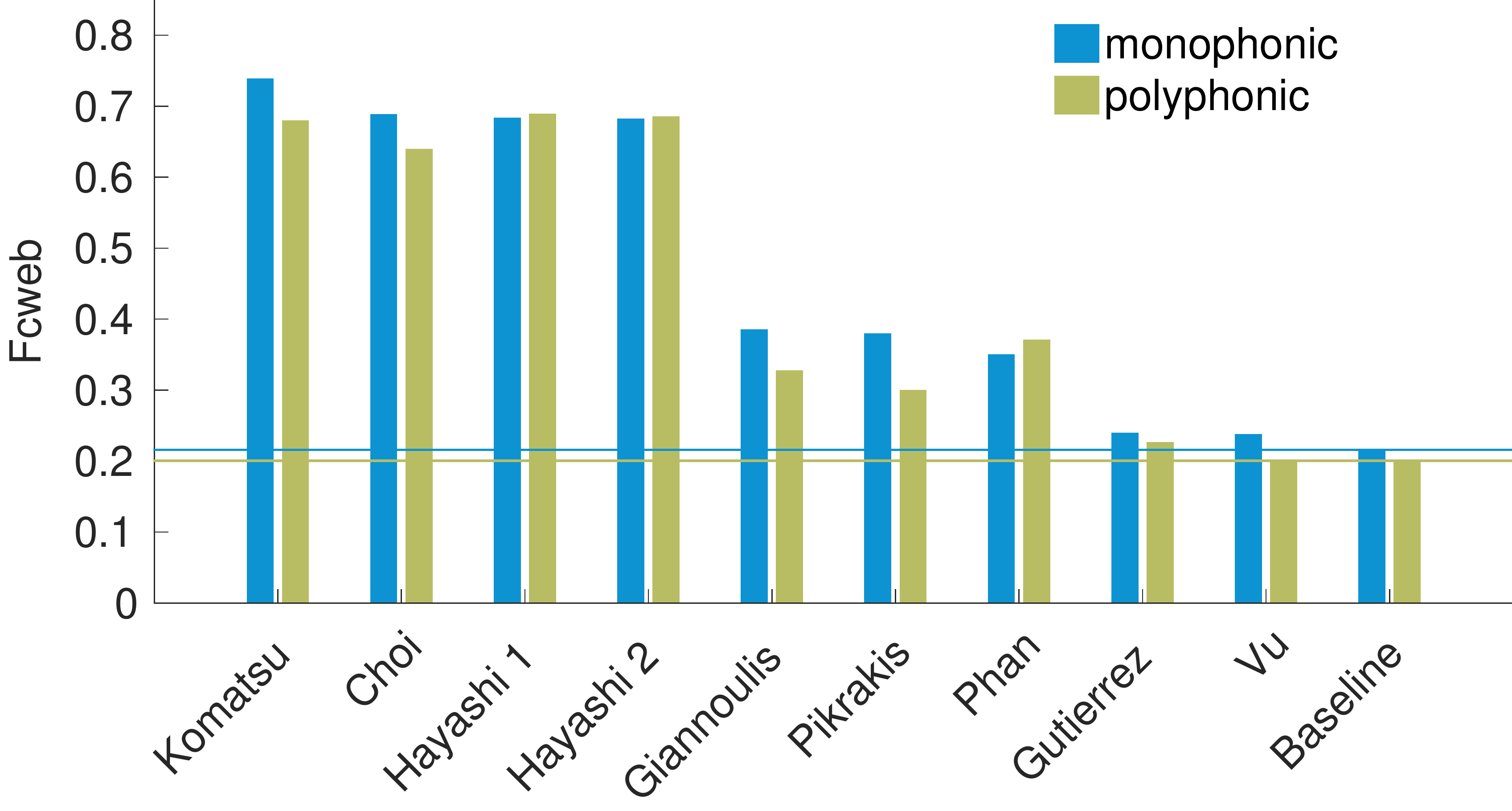}
\vspace{-0.15in}
\caption{Impact of polyphony on the performance of the systems submitted to DCASE 2016 - Task 2, in terms of $F_{\textit{cweb}}$. Baseline performance is shown with a straight line.}
\label{fig:overall_poly_eb_class_wise_F}
\end{figure}

Results by scene type (monophonic vs. polyphonic scenes) are plotted on Fig.~\ref{fig:overall_poly_eb_class_wise_F}. The ANOVA analysis performed on $F_{\mathit{cweb}}$ shows a positive effect of the type of system ($F[9,468]=358$, $p_{gg}<0.01$), but no significant effect of polyphony ($F[1,52]=3.5$, $p=0.07$). An interaction between scene type and system is nevertheless noted.

Thus, perhaps surprisingly, the scene type does not affect system performance, as the systems are on average able to handle both scene types equivalently. \emph{Post hoc} analysis on the scene type shows that among the 10 systems, 4 have their performance reduced while considering polyphonic scenes: \emph{Choi}, \emph{Giannoulis}, \emph{Komatsu} and \emph{Pikrakis}. For those systems, performance is reduced for polyphonic scenes, which probably explains the significant effect of interaction between scene type and system showed by the ANOVA analysis.

\emph{Post hoc} analysis on the type of system considering monophonic or polyphonic scenes allows us to cluster submitted systems into the same 3 groups as those identified with the global performance, with the \emph{Kong} system being removed. The only difference is the loss of the significant difference between systems \emph{Gutierrez} and \emph{Pikrakis} on polyphonic scenes.

\subsection{Background Level}

\begin{figure}[t]
        \subfloat[]
        {\includegraphics[width=1\linewidth]{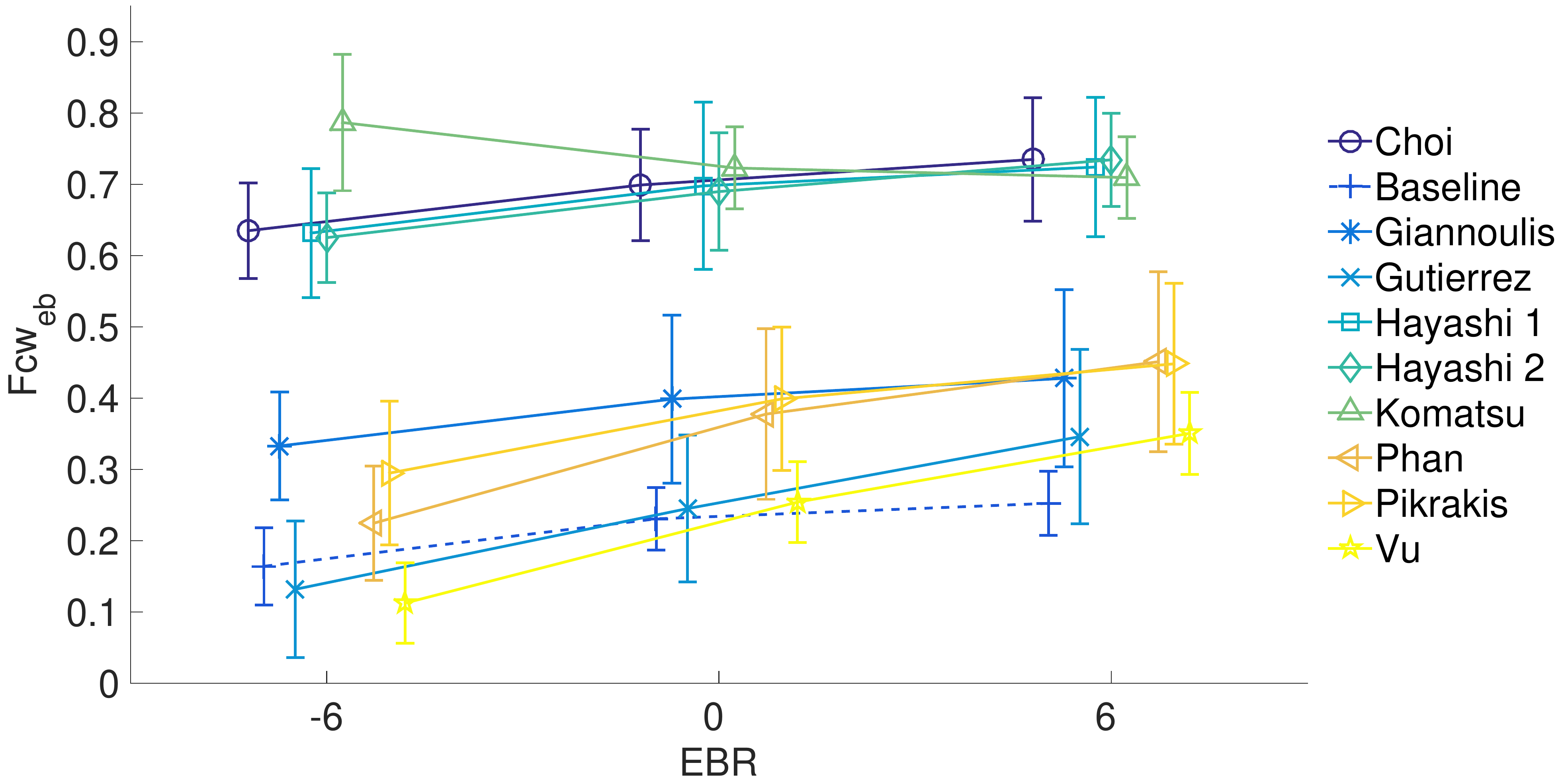}\label{fig:results_ebr_poly0_eb_class_wise_F}}\vspace{-0.15in}\par
        \subfloat[]
        {\includegraphics[width=1\linewidth]{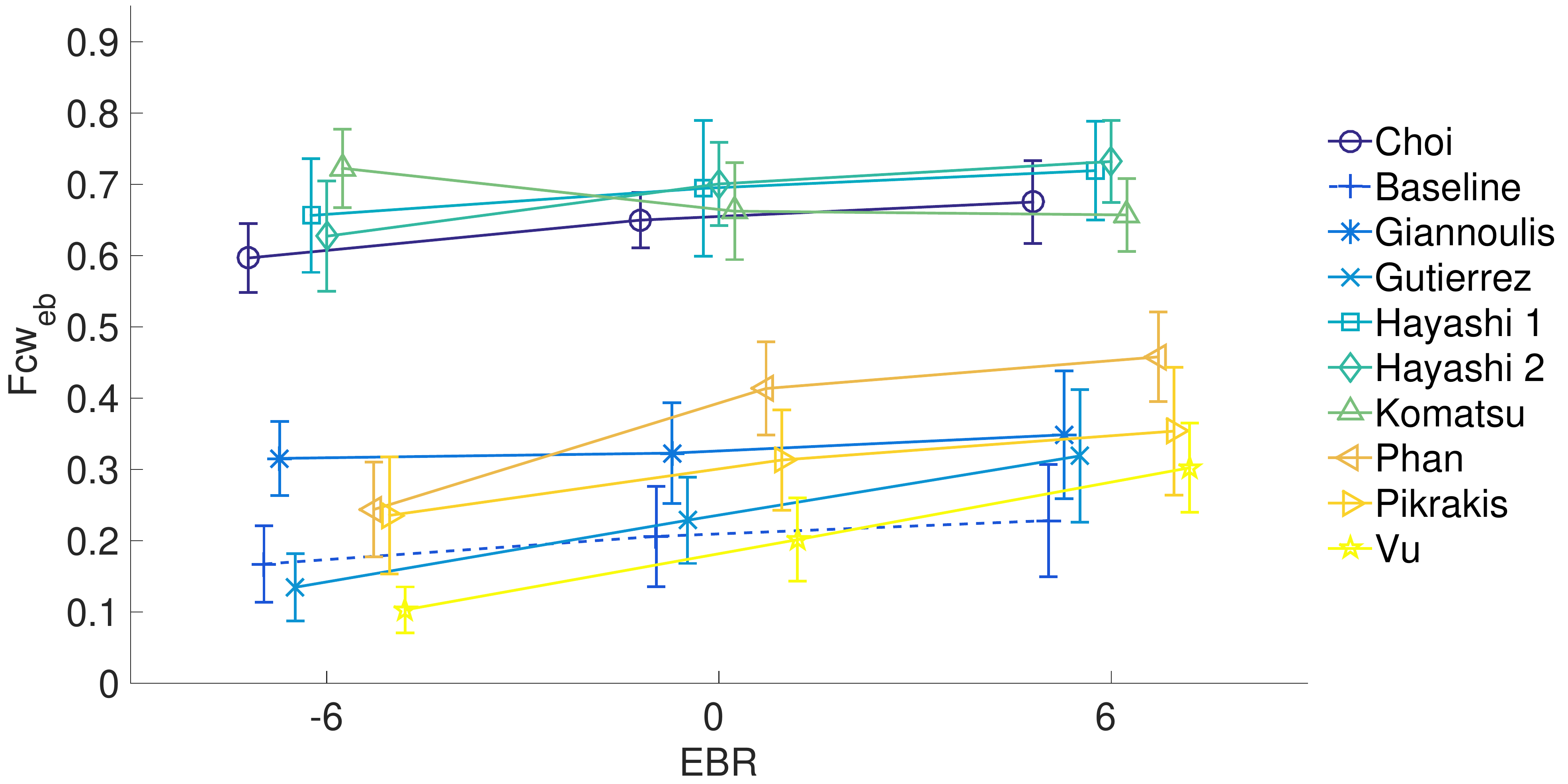}\label{fig:results_ebr_poly1_eb_class_wise_F}}\par
       \caption{Impact of the background level ($EBR$) on the performance of the systems submitted to DCASE 2016 - Task 2, considering  $F_{\textit{cweb}}$ for (a) monophonic and (b) polyphonic scenes.}\label{fig:dcase2016_poly1_eb_fc}
\end{figure}

Results on varying levels of background noise are plotted on Fig.~\ref{fig:results_ebr_poly0_eb_class_wise_F} for monophonic scenes. The ANOVA analysis shows a significant effect of the type of system ($F[9,72]=80$, $p_{gg}<0.01$), the $EBR$ ($F[2,16]=164$, $p_{gg}<0.01$), and their interaction
($F[18,144]=6.5$, $p_{gg}<0.01$). This interaction shows that the higher the $EBR$, the better the system performs, and this stands for all the systems except one: \emph{Komatsu}.

For the  \emph{Post hoc} analysis, we study if the submitted systems reach performance that is significantly superior to the one achieved by the baseline.
To summarize the results for the monophonic scenes, the \emph{Komatsu} system achieves the best performance, particularly for high levels of background ($EBR=-6dB$). This is the only system that has its performance decreasing as a function of the $EBR$, probably due to the proposed noise estimation dictionary learned during the test phase.
The systems of  \emph{Choi}, \emph{Hayashi 1} and \emph{Hayashi 2} systematically perform better than the others and equal the \emph{Komatsu} system for $EBR$ of $0$ and $+6dB$. For those 3 systems, raising the background level ($6dB\rightarrow -6dB$) leads to a drop of performance of about $10\%$. The remaining systems outperform the baseline only for some levels of $EBR$. Those systems appear not to handle robustly the background, as their performance is notably lower as a function of its level, from  $-10$ to $-20\%$ for an $EBR$ of $6dB$ to $-6dB$. Only \emph{Gutierrez} maintains its superiority with respect to the baseline in all cases.

Considering polyphonic scenes, results are plotted on Fig.~\ref{fig:results_ebr_poly1_eb_class_wise_F}. The ANOVA analysis shows a significant effect of the type of system ($F[9,72]=113$, $p_{gg}<0.01$), of the $EBR$ ($F[2,16]=127$, $p_{gg}<0.01$), and of their interaction ($F[18,144]=15$, $p_{gg}<0.01$). As with the monophonic scenes, the higher the $EBR$, the better the system performs, and this stands for all systems except one: \emph{Komatsu}.
%
%
%
%
%
Results for polyphonic scenes are thus similar to the ones achieved on monophonic scenes at the exception of 1) \emph{Vu}, \emph{Pikrakis} and \emph{Gutierrez} that have performance equivalent to that of the baseline for all $EBR$ levels and 2) \emph{Phan}, which reaches performance that is above the baseline for $EBR$s of $0$ and $+6dB$.

\subsection{Number of Events}

\begin{figure}[t]
        \subfloat[]
        {\includegraphics[width=1\linewidth]{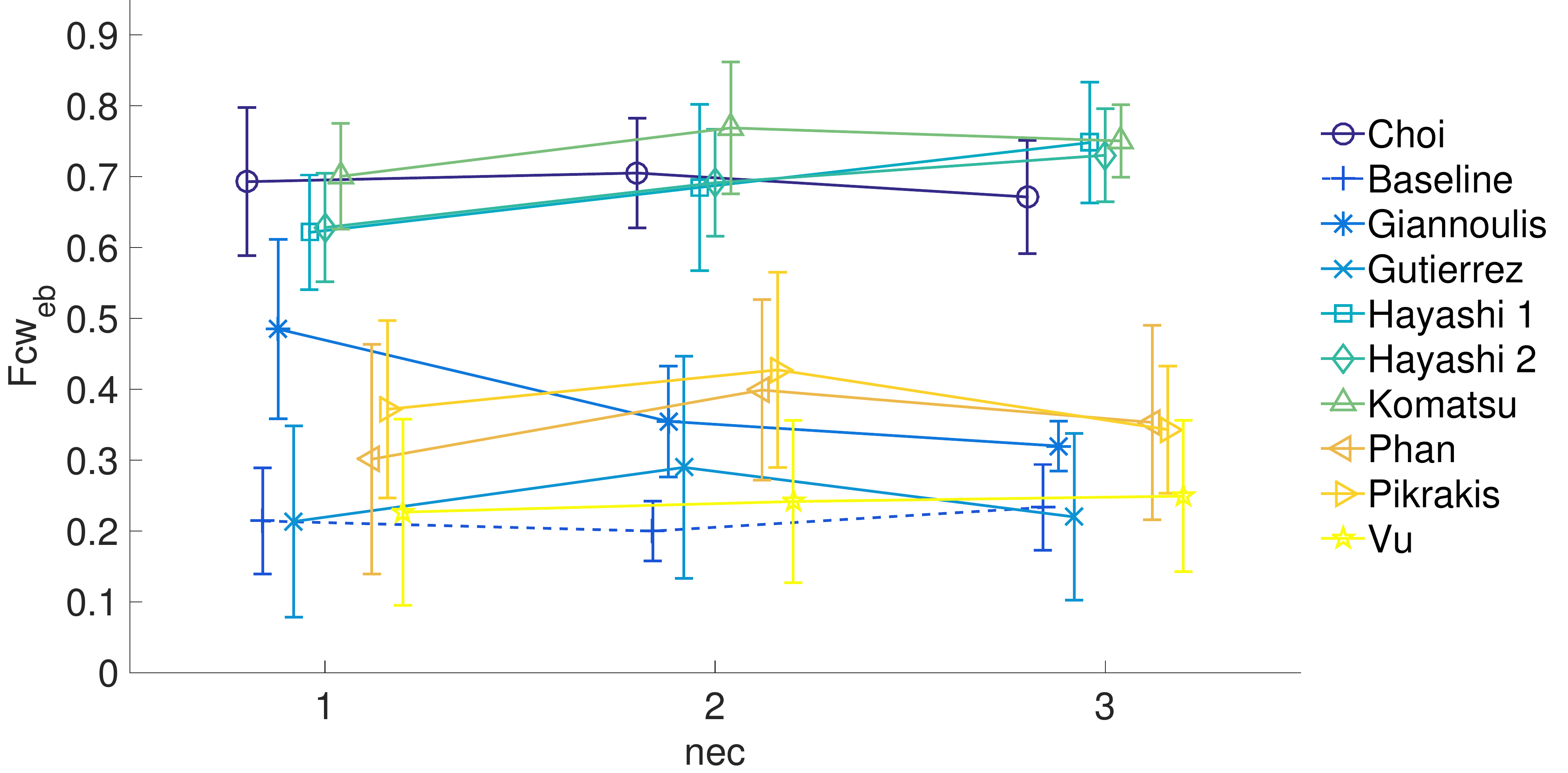}\label{fig:results_dens_poly0_eb_class_wise_F}}\vspace{-0.15in}\par
        \subfloat[]
        {\includegraphics[width=1\linewidth]{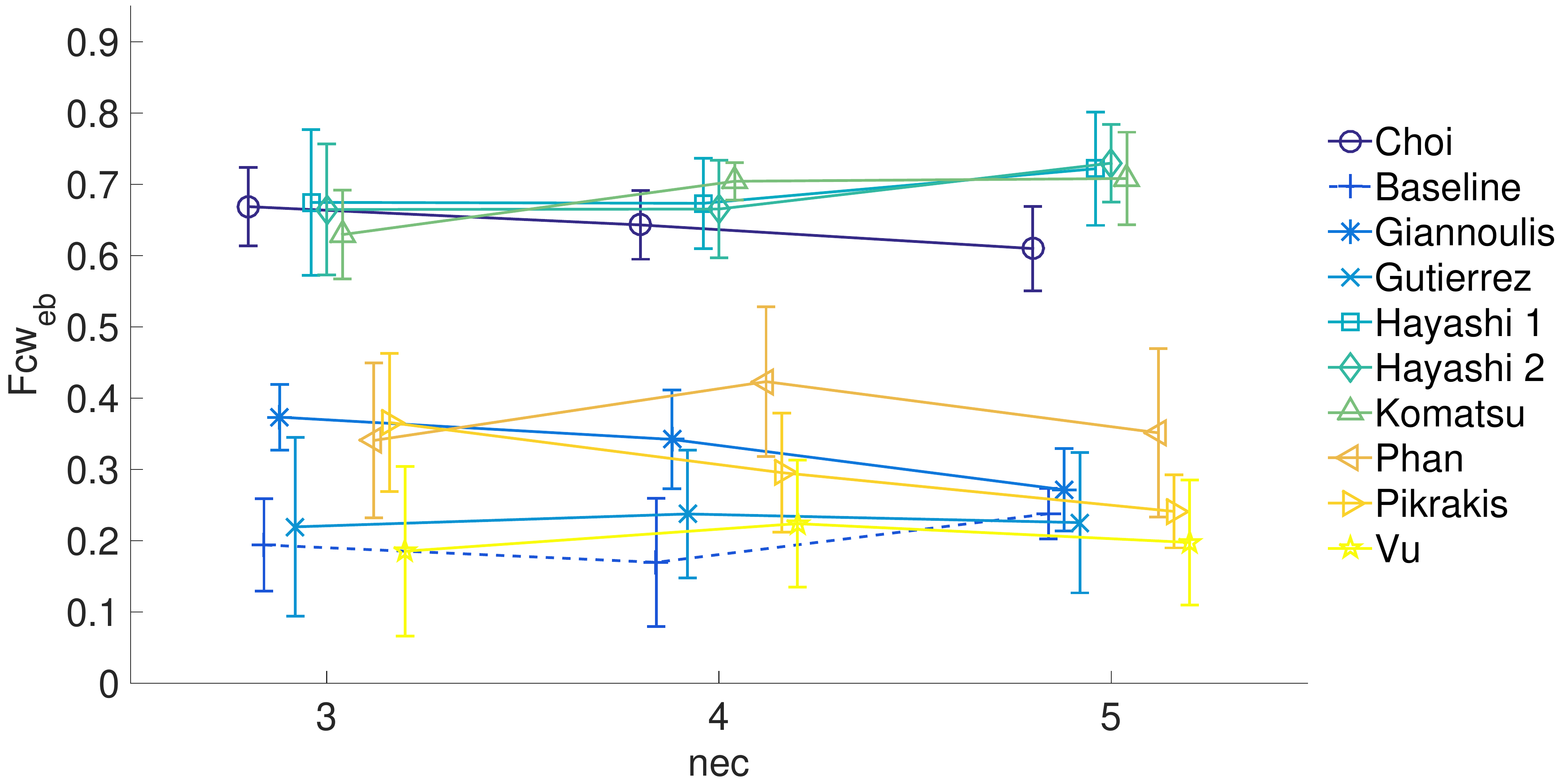}\label{fig:results_dens_poly1_eb_class_wise_F}}\par
        \vspace{-0.05in}
       \caption{Impact of the number of events ($nec$) on the performance of the systems submitted to DCASE 2016 - Task 2, considering $F_{\textit{cweb}}$ for (a) monophonic and (b) polyphonic scenes.}\label{fig:results_dens_eb_class_wise_F}
\end{figure}

Considering monophonic scenes, results are plotted in Fig.~\ref{fig:results_dens_poly0_eb_class_wise_F}. The ANOVA analysis shows a significant effect of the type of system ($F[9,216]=264$, $p_{gg}<0.01$), but not on the number of events ($nec$) ($F[2,24]=0.5$, $p=0.6$). An interaction between the two can nevertheless be observed.
Similar conclusions are obtained for polyphonic scenes (Fig.~\ref{fig:results_dens_poly1_eb_class_wise_F}). The ANOVA analysis shows a significant effect of the type of system ($F[9,216]=170$, $p_{gg}<0.01$), but not of $nec$ ($F[2,24]=0.1$, $p=0.9$) and some interaction between the two.
Consequently, it is difficult to conclude on the influence of $nec$ on the differences between systems.

Some trends can be observed though. For monophonic scenes, augmentation of the number of events leads to a systematic performance improvement of 2 systems
(\emph{Hayashi 1}, \emph{Hayashi 2})
and a decrease for 1 system (\emph{Giannoulis}).
For polyphonic scenes, an improvement is observed for 3 systems
(\emph{Hayashi 1}, \emph{Hayashi 2}, \emph{Komatsu})
and a decrease for 3 others
(\emph{Giannoulis}, \emph{Pikrakis}, \emph{Choi}).
Contrary to the level of background which on average reduces the performance of the systems, it appears that the impact of the number of events varies from system to system. 
\emph{Hayashi 1} and \emph{Hayashi 2} react positively to the augmentation of $nec$ whereas \emph{Giuliano} has its performance systematically decreasing, probably due to different miss detection / false alarm tradeoffs.

\section{Discussion} \label{sec:discussion}

Among the 3 experimental factors evaluated, only the $EBR$ seems to have a significant influence on the performance of the algorithm. No significant impact is observed for the polyphony and the number of events. Those results clearly demonstrate the usefulness of considering carefully the background in the design of the system.

For all the $EBR$s considered, 4 systems (\emph{Hayashi 1}, \emph{Hayashi 2}, \emph{Komatsu} and \emph{Choi}) significantly have better performance than the baseline. Among those, \emph{Komatsu} is the only one that obtains significantly higher performance than the other systems, this for an $EBR$ of $-6dB$. Thus, the \emph{Komatsu} system can be considered as the system with the best generalization capabilities, thanks to an efficient modelling of the background.

In light of those results, we believe that considering simulated acoustic scenes is very useful for acquiring knowledge about the properties and the behaviors of the evaluated systems in a wider variety of conditions that can be tested using recorded and annotated material.
We acknowledge that the sole use of simulated data cannot be considered for definitive ranking of systems. Though, considering only recorded and manually annotated data as of today's scale does not allow researchers to precisely quantify which aspect of complexity of the scene is impacting the relevant issues to tackle while designing recognition systems. Recorded data are indeed often scarce resources as the design of large datasets that have a wide variety of recording conditions is complex and costly. Also, reaching a consensus between annotators can be hard.


Taking a more methodological view, such a use of simulated data has recently received a lot of attention in machine learning research, inspired by experimental paradigms that are commonly used in experimental psychology and neuroscience where a special care of the stimuli is taken in order to inspect a given property of the system under scrutiny.
This principle led to some interesting outcomes in a slightly different paradigm for a better understanding of the inner behavior of deep learning architectures \cite{Goodfellow15, Nguyen15}.

\bibliographystyle{IEEEtran}
\bibliography{bibliography.bib}

\end{sloppy}
\end{document}